# Universal Model for Ferroelectric Capacitors Operating Down to Deep Cryogenic Temperatures

*Ella Paasio, Rikhard Ranta, Sayani Majumdar\**

Faculty of Information Technology and Communication Sciences, Tampere University, 33720 Tampere, Finland

*E-mail: sayani.majumdar@tuni.fi

Keywords: non-volatile memory, analog memory, ferroelectric capacitors, ferroelectric $Hf_{0.5}Zr_{0.5}O_2$, Jiles-Atherton Model, Neuromorphic, Quantum

Binary oxide ferroelectrics like doped $HfO_2$, compatible with complementary metal-oxide-semiconductor (CMOS) platforms, have gained significant interest for energy efficient, scalable, high-performance non-volatile memory and neuromorphic technologies. However, there is a gap in models for doped hafnia systems that can explain physical properties while being circuit simulation compatible and computationally efficient. We present a universal model based on the Jiles-Atherton equations to reproduce experimentally measured polarization switching in ferroelectric thin film capacitors under different electric field and temperature conditions. Additionally, device-to-device variation effect on the model parameters is presented, which will enable large-scale integration of the FE components to complex functional circuits. Due to increased interest in cryogenic electronics for quantum computing and space technologies, effect of temperatures on polarization switching is investigated down to 4 K. We show our model can reproduce the experimental polarization-voltage relation of Hafnium Zirconium Oxide (HZO) capacitors with nearly 100 % accuracy, for different electric fields and temperatures down to 4 K, including analog switching. We find cooling the devices below 100 K increases polarization update linearity and symmetry significantly. Our results represent an important advancement towards application of ferroelectric HZO capacitors for large-scale memory and neuromorphic circuits operating down to deep cryogenic temperatures.

## 1. Introduction

Current computing technologies rely on designing and precision engineering of billions, densely packed atomic-scale on-chip transistors to perform logic and memory operations. Lateral size of the logic transistors has reduced following the Moore's law since the 1950s. This lateral scaling required vertical scaling of the gate oxide materials to maintain the field effect on the channel. Prior to the release of the 45-nm node around 2007, the then-state-of-the-art planar transistors could include a gate oxide layer of $SiO_2$ thinner than one nanometer, equivalent to about three monolayers. In these ultrathin films, thickness dependent quantum mechanical tunneling effects start to dominate, leading to high gate leakage current, which significantly increases the off-state power consumption of the devices. To circumvent this problem, high-$k$ dielectric oxides were introduced in the gate stack. Implementation of the 45-nanometer node by Intel marked this pivotal shift from $SiO_2$ to hafnium oxide ($HfO_2$) as the dielectric material. [1]

Complementary metal oxide semiconductor (CMOS) compatible memories like static random-access memory (SRAM), dynamic random-access memory (DRAM) and flash technologies are the current mainstream memories. Scaling of these memory technologies has not been able to keep up with that of the logic transistors, producing a significant gap. Additionally, volatile data retention of SRAM and DRAM, and high voltage requirement for flash have remained a challenge. In recent years, multiple emerging technologies have shown promises for non-volatile data retention, including resistive random-access memories (RRAMs), [2] phase change memories (PCRAMs), [3] magnetic random-access memories, [4] among others. From the energy-efficiency, fast operation and stable data retention perspective, highly scalable ferroelectric (FE) memory devices [5,6] provide unparalleled advantages over RRAMs and PCRAMs. Additionally, CMOS compatibility, high thermal stability, forming free operation, and matured fabrication process technologies like atomic layer deposition (ALD) make the FE memory devices attractive for industry scale applications. Large polarization switching in FE hafnia-based systems offers a wide conductance window and high bit precision in FE field effect transistors (FeFETs), thanks to their stable domain rotation properties. These make FE memories highly suitable for online learning applications in neuromorphic computing.

FE hafnia-based devices are well suited for in-memory computing (IMC) and neuromorphic spiking neural network (SNN) applications, for which the intriguing and intricate physics in the electrical behavior can be utilized for both neural and synaptic devices. [5] A simple leaky integrate-and-fire (LIF) neuron model can be implemented with harnessing the accumulative switching behavior in either FTJs [6] or as simply as a two-transistor circuit consisting of FeFETs. [7,8] The intrinsic stochasticity of the domain formation and reversal in polycrystalline FE can be beneficial for neuromorphic applications requiring a level of stochasticity to follow brain-inspired learning rules. These applications are mostly for low-energy adequate-accuracy applications, where prediction accuracy can be lowered to reach the major improvement in energy efficiency. Software gradient descent based DNN models can reach higher accuracies, but in various applications these incredibly high accuracies are not worth the enormous energy consumption. [5]

In recent years, FE Hafnium Zirconium oxide (HZO) based 1-transistor – 1-capacitor (1T-1C) FE random-access memory (FeRAM), FeFETs and FTJs have become a subject of significant research interest due to their compatibility with CMOS back-end-of-line (BEOL) processes. HZO-based systems are interesting not only for non-volatile memory (NVM) and as synaptic weight elements for neuromorphic computing and IMC [9,10] but also for other applications such as sensing [11,12] and radio-frequency (RF) electronics. [13] The complexity of low-thermal budget fluorite FE structures, arising from coexistence of multiple phases and structural defects due to polycrystallinity, makes understanding of the physics and engineering of HZO based systems more challenging compared to that of traditional single crystal perovskite ferroelectrics. [14] Added complexity from random domain formation and rotation due to inhomogeneous distribution of defects and domain pinning sites in HZO thin films causes certain stochasticity in switching and wafer-scale inhomogeneity in the devices. Therefore, a proper understanding of the material switching properties, and a physics-based modelling of FE device performance is crucial, not only for further understanding of the associated physical phenomena, but also to simulate circuit level operation of FE devices.

A novel research direction for HZO-based FE devices that is gaining momentum is its cryogenic operation, as cryogenic memories are becoming crucial components for the quantum information processing. [15] In addition to quantum technologies, large and fast memory systems for high performance computing (HPC), space technologies and superconducting electronics are other interesting possible applications of cryogenic circuits with FE components. [16] There have been recent research results showing stable [17,18] and even improved [19,20] FE behavior at cryogenic temperatures. This work focus on modelling the HZO devices, reported in [19] down to 4 K for their use in cryogenic electronics.

Several simulation models have been used for modelling the FE polarization switching, which for hafnium-based devices most often fall into categories of Landau-based, [21–23] Preisach-based [24–27] and nucleation limited switching (NLS)-based. [28–30] However, their parameters either lack analytical physics-based definitions due to their empirical nature, [24, 27–30] or are computationally very demanding [25,26] and thus incompatible with circuit simulation software. Adaption of the well-established ferromagnetic Jiles-Atherton (JA) model [31] has been used in literature to model polarization switching in FE devices, [32–35] and the SPICE and Verilog-A compatibility of the model has been demonstrated. [32] In the current work, we investigate the applicability of the JA model to thin-film HZO capacitors as function of applied electric fields at temperatures between 300 K to 4 K and show the universality of the JA model to accurately reproduce the device behavior over a broad $E$ and temperature range. The model was validated using experimental polarization-voltage ($P$-$V$) and current-voltage ($I$-$V$) hysteresis curves. [19] Our modelling results highlight the suitability of the JA model for reproducing the major and minor loop operations in polycrystalline FE thin film devices over a wide temperature range. With the extracted model parameters, we can successfully predict the analog nature of switching, consistent with results predicted from material physics and experiments. We analyze the results further through Hamiltonian Monte Carlo simulations, which affirm the simple physical explanation for the discoveries. These findings can provide important design guidelines for large-scale integration of non-volatile, analog memory devices based on HZO thin films over a temperature range of 4 – 300 K and hold promises for acceleration of their integration with CMOS platforms for a broad range of applications, starting from standard classical computing, neuromorphic edge computing to cryogenic quantum computing and space technologies.

## 2. Results and Discussion
### 2.1. Ferroelectric hysteresis modelling with Jiles-Atherton equations

Schematics of the FE capacitor stack, used in this work, is shown in **Figure 1 a)**. Details of the experimental techniques, both fabrication and characterization used for the model validation, is described in our previous work [19] and in the Methods section for ease of access. The JA model considers the polarization on a domain level and is based on the kinetics of the domain walls. [33] Domains, which are groups of neighboring unit cells with similar dipole moments with local translational symmetry, are shown in **Figure 1 b)**, and are separated from each other through domain walls (**Figure 1 c)**. The driving force for the switching of polarization states in the JA model is a pressure experienced by the domain walls due to a difference in the thermodynamical potentials between the two sides of the domain wall. [36] In a thin film capacitor, where the depolarization field contributes significantly to the remanent polarization ($P_r$) of the material due to finite electron screening lengths in the electrodes, [37,38] the depolarization field is included in the total electric field ($E$) in the capacitor. In the JA model, the ideal (anhysteretic) polarization is shaped like the Langevin equation, [33] obtained from assuming that the rotation of dipole moments is not restricted, and instead the dipole moments can rotate in all spatial directions when an external electric field is present. The Langevin equation is comparable to the mean field approximation of the well-known Ising Hamiltonian, where it is assumed that polarization is either fully up or down, with no intermediate states. [33] The microscopic anhysteretic polarization is scaled into the macroscopic description level by the saturation polarization parameter, $P_s$, which describes the ideal total polarization without hysteretic losses. Another shape parameter for the anhysteretic polarization is known as the domain wall density parameter, $a$. In fluorite FEs, the domain walls are very thin, mostly corresponding to a wall of one- or two-unit cell thick walls separating two domains, [39] so definition of the parameter is in need of modification in this system. In analogy to the domain wall density in ferromagnetic materials, that corresponds to the effective susceptibility of the polarization reversal, [33] the parameter $a$ can be viewed to encompass an inverse of domain wall velocity in the case of fluorite FEs.

Markedly, the polarization in the JA model is purely FE polarization, thus corresponding to measurement data obtained from positive-up-negative-down (PUND) measurements, where the dielectric and leakage contribution have been subtracted from the total current data to obtain purely FE contribution. The JA model is based on the kinetics of multiple domains propagating in the material, and an important characteristic of the kinetics is the strength of dipole interactions with each other, which is encompassed by the inter-domain coupling term, $\alpha$. Notably, the dipole and subsequent domain interactions are influenced by external applied stress, and the parameter $\alpha$ is in addition to the effective dipole charges and distances, also dependent on applied strain. [33] Interactions with pinning sites, such as oxygen vacancies, defects or other interstitials also impede the domain wall propagation. These pinning sites account for two distinctive polarization switching modes of domain wall kinetics. First, the domain walls can get pinned on these pinning sites, where the domain wall will bulge around the pinning site, as the domain wall energy is insufficient to de-pin the domain. Secondly, if the energy of the domain wall is sufficient to break the pinning site, the domain can de-pin, irreversibly moving the pinning site. The total loss in energy caused by breaking the pinning sites on a macroscopic level is encompassed into the pinning loss parameter, $k$, which is dependent on both the average energy to break the pinning sites, and the number of sites to be broken.

The total contribution to the polarization by breaking pinning sites corresponds to the irreversible polarization, and the change in polarization due to the bending of the domain walls without breaking the pinning sites corresponds to the reversible polarization. Using these definitions, a differential equation for the irreversible polarization can be obtained and solved by using the simple Euler integration method, as shown in Eq. (9). In total, the polarization corresponds to both reversible and irreversible polarization, with the polarization reversibility factor, $c$, describing an effective ratio of the contributions. This set of equations (Eqn. 1 - 9) provides a physics-based framework to model the polarization switching in FE materials, including polycrystalline HZO. The equations and the integration procedure are shown in full in the Methods. The integration procedure more commonly used with the JA model are finite element methods since they provide better convergence in ferromagnetic hysteresis. [40] However, for FE thin films that are one-dimensional problems due to measured current being one-dimensional, the computational efficiency of the Euler method outweighs the possible stability finite element methods could provide, and convergence does not pose an issue for the simulations in this work.

The JA model is in many ways comparable to the Preisach model [44] based on the assumption that the material consists of simple hysteresis-wise square-like hysterons, that have coercive fields ($E_C$) from a distribution of possible fields, solved through a computationally demanding double integral. In addition to the Preisach model and approximations of it, commonly used for their convenience in producing the $P-V$ curve, the NLS model [43] has been well established in HfO$_2$-based FE materials. The model has been built upon the assumption that in thin films nucleation events limit the switching at small switching time constants and low $E$. This is directly contradicting the Kolmogorov-Avrami-Ishibashi (KAI) model, where after nucleation the domain growth is unrestricted and always dominating. [43] The NLS switching times follow the semiempirical Merz law, [44] which states that the switching time constants depend exponentially on an effective activation field, connected to domain wall depinning [38] and temperature, [45] and the domain wall movement is nearly restricted into the direction parallel to the electric field. [44]

Notably, the results obtained with the NLS model, and the physical basis of the JA model do not mismatch, but for signals consisting of sub-microsecond pulses with low voltages, the NLS model can better encapsulate the switching dominated by nucleation. The complete NLS model is best suited for Monte Carlo simulations, [46] further adapted into a SPICE compatible format based on two integration processes, a sampled distribution, as well as a time constant update protocol. [47] The algorithm for solving the polarization with the JA model is more effective to implement and able to reproduce the polarization switching behavior accurately, when the energy added to the system with the bias voltage is high enough for domain wall kinetics to dominate over nucleation.

The JA modelling provides a way to directly differentiate the domain wall movement dependent phenomenon from other device level phenomena, such as fatigue or wake-up, by analyzing the PUND data and comparing it to the theoretically constructed hysteresis curve. Beside the deep understanding of the physics governing device performance, modelling the polarization dependence on the input voltage is an important step towards modelling the (*I-V*) properties of the FE devices, upon which all simulation-based circuit design is based. The displacement current induced by the polarization switching is simply the time derivative of the polarization, and other device behavior like dielectric contribution, leakage path current, or tunneling current in FTJs modify the *I-V* characteristics beyond this simple derivative approach. However, polarization is the baseline to understand all the other phenomena, and by computationally modelling the polarization switching over a wide $E$ and temperature ($T$) range, we aim to accelerate the development of efficient and accurate circuit simulation models, while providing deeper understanding of the device operation on a semi-microscopic level.

The transient currents in response to triangular voltage pulses, measured from our $Hf_{0.5}Zr_{0.5}O_2$ (HZO) capacitor, are shown in **Figure 1(d)**. The experimental results, collected by PUND technique between 300 K – 4 K, with a range of amplitudes of voltage pulses to study the effect of *E* and *T* on the model parameters. Transient *I-V* and *P-V* loops are computed from the experimental data for multiple amplitudes of peak-to-peak voltages ($V_{pp}$). As shown in **Figure 1(e)**, the experimental *P-V* curve, measured at 10 $V_{pp}$ pulse amplitude and 500 Hz frequency, is a representative example that the modelled hysteresis shows excellent agreement with experimental data when comparing the remanent polarization $P_r$, coercive fields $E_{C+}$, $E_{C-}$ and area of the *P-V* loop.

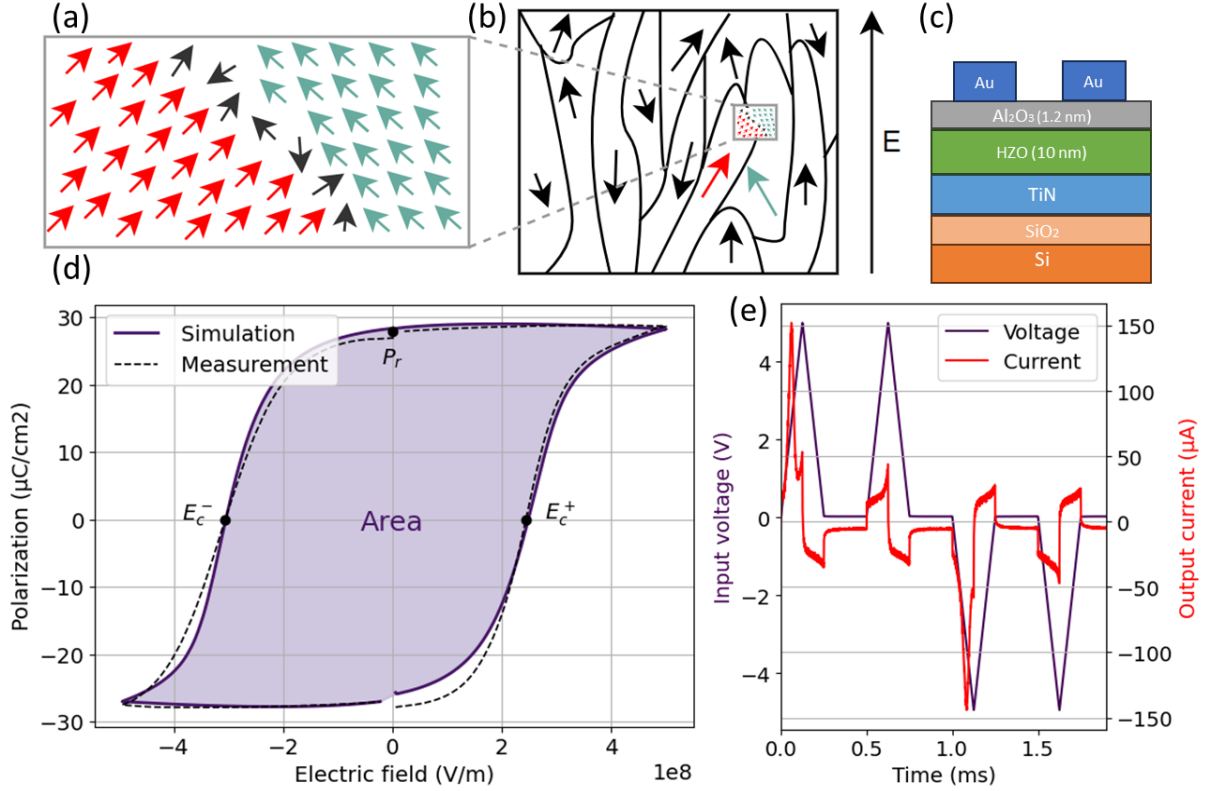

**Figure 1.** (a) Macroscopic ferroelectric polarization consists of microscopic unit cell sized dipole moments, represented by the arrows. (b) The dipoles forming domains separated by domain walls, the propagation of which is mostly in line with the direction of the electric field (c) HZO capacitor stack used in the experiment and simulation. (d) PUND voltage cycling with 500 Hz and 10 $V_{pp}$ and corresponding current output. (e) A simulation result with the Jiles-Atherton model where the parameter values are : $a$ = 4.16 C/m$^2$ , $\alpha$ = 0.6345, $k$ = 68.66 C/m$^2$, $c$ = 0.6980, and $P_S$ = 29.63 C/m$^2$.

To gain insights on the physical origin of the model parameters and their impact on the shape of the *P-E* hysteresis as well as on the dynamic *I-V* curve, we varied one parameter value at a time while keeping the other parameters constant (**Figure 2**). Here, a positive voltage pulse is applied first followed by a negative one, while the reverse scenario is shown in the supporting information (**Figure S1**). The parameter *a*, which in other implementations is either referred to as a shape parameter [32, 33] or as domain wall density [34] shows a trend that an increase in *a* leads to a reduction in the slope of the curve at the turning points. Additionally, the onset of switching occurs at a lower *E* when *a* is increased. The impact of *a* is symmetric, with both pulse schemes (positive first or negative first) producing identical hysteresis curves. From this analysis, and the derivation by Smith et. al, [33] where *a* is connected to the permittivity of the material, we can deduce that *a* is connected to the velocity of the domain walls in the material, which is lowered by properties such as domain wall density. The parameter, *α*, predicts how strongly the domains are coupled. As the polarization is initialized to one side, strong coupling leads to a built-in bias field prohibiting the domain rotation. An increase in α forms a wider hysteresis curve, as shown in **Figure 2**, affecting $E_C$ values on each side, while also influencing the slope of the curve during switching. Depending on the initial polarization direction, one polarization direction is more preferred compared to the other. Effect of varying pinning loss parameter, *k*, is shown in the third column of **Figure 2**. A higher value of *k* indicates an increased number of low energy defects where de-pinning is possible. This de-pinning consumes energy, restricting the total energy for the domain wall movement. When positive voltage is applied first, the transition from positive to negative *E* does not significantly affect the *P-E* loop with changing *k*. However, for reverse switching, a higher *k* reduces the magnitude of *E* required, causing narrowing and asymmetry in the *P-E* hysteresis. Pinning loss *k* has an even more significant asymmetric nature compared to the parameter, α. Irreversibly moving the pinning sites to one side enforces the built-in bias field, favoring switching more in one direction than the other. The polarization reversal parameter, *c*, shows with increasing *c*, a more saturated hysteresis curve can be formed, while a lower value of *c* leads to inward bending of the curve at the turning points. A higher polarization reversibility indicates a higher number of pinning sites with high enough energy to stay pinned during field cycling. Increasing *c* slightly increases the $P_r$. The parameter, $P_S$ denotes the maximum value of anhysteretic polarization, which would be the total polarization in an ideal material with unimpeded domain movement. The parameter value also influences the $E_C$ s on both sides, so it has a major effect on the area under the curve.

As shown from the modification of the memory window in the inter-domain coupling and pinning loss parameters in **Figure 2,** the JA model is directly capable of modelling devices that shows asymmetric switching on the *E* axis, due to asymmetry in stack configuration or the imprint effect**.** While asymmetry in switching due to the imprint effect is in some memory applications viewed as a detriment to device performance, [48,49] for brain-inspired computing it could be harnessed as a feature instead of a flaw. [50] The simulation is, in addition to the value of the parameters, found to depend on whether a positive or a negative pulse was used to initialize the model. This initialization effect is seen most prominently in the pinning loss parameter, *k*, as well as the inter-domain coupling, *α*. Rest of the parameters are not dependent on the direction of the initialization of the device in the simulation. The imprint observed with the inter-domain coupling can be explained by the interplay between the *E* pursuing to turn the polarization state, and the coupling holding the domains in their original state. A higher coupling term indicates that a stronger internal bias field is formed, and a higher $E_C$ is required to flip the polarization from its once initial state consisting of strongly coupled domains. With a lower inter-domain coupling, the bonds holding domains aligned is weaker, making them more prone to reversal. For the pinning loss parameter k, this effect originates mostly from pinning sites irreversibly breaking during the initial pulse, making one direction preferred over the other. A higher loss can indicate a higher number of mobile pinning sites caused by, for instance, trapped charges at oxygen vacancies in the HZO device stack, a commonly referred explanation for imprint effect, [48,51] which is agreement with the behavior seen in **Figure 2**.

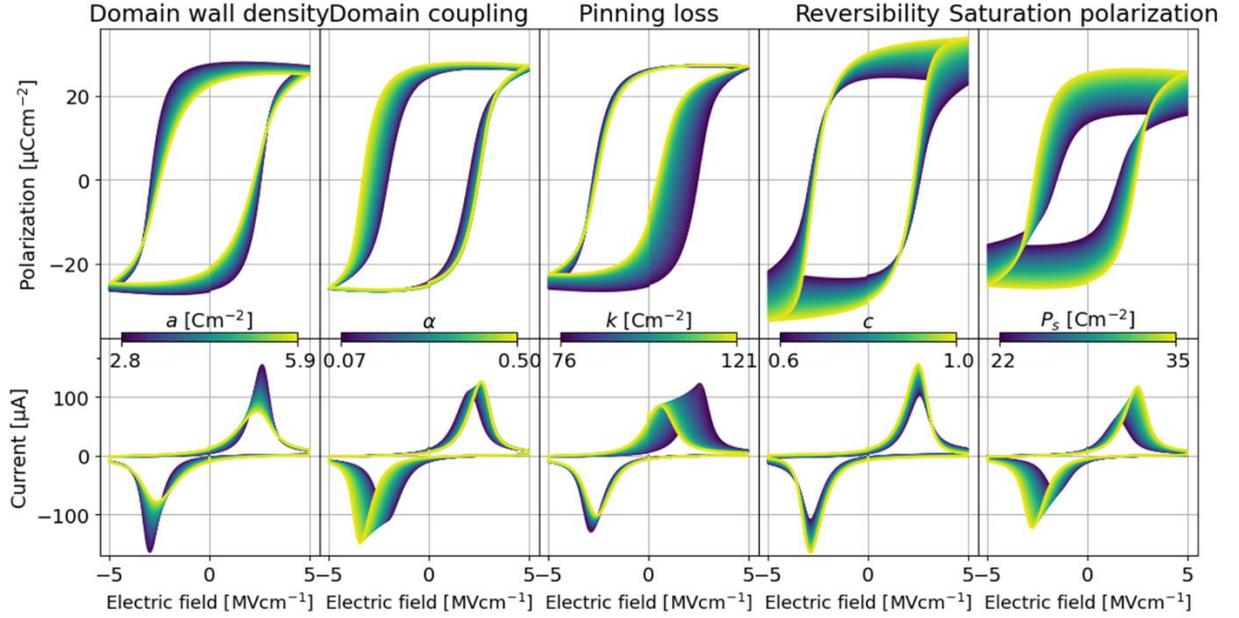

**Figure 2**: Changes in the hysteretic *P-E* curves with modifying one JA model parameter while keeping others constant and the corresponding *I-E* curves. The color indicates the value of the modified parameter, and the asymmetry in the case of inter-domain coupling and pinning loss parameters is dependent on the initial pulse direction.

## 2.2. Modelling device-to-device variation

For a realistic modelling of HZO devices in large system integrations, extrapolating the characteristics of just one device for an array of devices without information of the device-to-device (D2D) variation may cause unrealistic and overfitting circuit simulation results. [52] Therefore, it is important to understand effect of inherent spatial variation on the model parameters. These wafer-scale spatial non-uniformities are connected to the inhomogeneous formation of polycrystalline phases due to the low thermal budget fabrication of the devices, leading to a distribution of $E_C$ and $P_r$ values. As the device technology is still in the developmental phase, D2D variation is larger than a fully automated manufacturing process would entail in near future. However, the spatial variation caused by the inherently stochastic properties of the low-temperature polycrystalline growth itself cannot be fully removed or ignored. One possible way to simulate D2D variation is by using Monte-Carlo methods with models such as NLS, [53] but this Monte-Carlo sampling does not provide SPICE compatible analytical equations. Other possible way to include variations are to sample device parameters from a statistically derived distribution. To effectively design circuits with more than one non-ideal FE device, we explore the statistics of the device characteristics as well as the simulation parameters.

The JA model does not inherently include stochasticity, so D2D variation needs to be accounted for by considering the variation of the model parameters from experimental curves. The $P - E$ curve measured from 73 separate devices fabricated on the same wafer are presented in **Figure 3 a)**. The $P - E$ curves exhibit a distribution of $E_C$ and $P_r$ values as well as areas enclosed by the curve, and the shape of the $P - E$ curve is provided in **Figure 3 b)**. The JA model is used to find the optimal parameters for each of the curves separately. The model accurately fits each of the devices, but due to differences in the values and shapes of the $P - E$ curves, the model parameters also vary between the devices. The distribution of the parameters is shown in **Figure 3 c)**, where the mean and standard deviation are highlighted. In addition, these values are used to assess error in the electric field dependency of the parameters. In **Figure 3 a)**, an average curve is computed by using the mean of each parameter value of this set of devices. This forms a representation of the average behavior exhibited by the devices, enabling construction of slightly differing curves for modelling desired D2D variation for large scale simulations.

The JA parameters provide semi-microscopic descriptions of the kinetics of domain walls in the material. In addition to statistical variation analysis, the parameters can also be plotted against each other, as is done in **Figure 3 c)**, to enable analysis of the parameters and their interconnectivity in terms of D2D variation. It not only provides an insight into the variation of the parameters, but also enables a deeper understanding of the dependencies these parameters have on each other. From a physical point of view, some parameter dependencies on each other can be explained. For instance, the pinning loss and polarization reversibility parameters ($k$ and $c$) have positive correlation, since both are related to the number of defects in the material but with different energies, and these parameters have a nearly linear dependency in **Figure 3 c)**. Some of the dependencies are not as clear, like the two clusters seen in the saturation polarization against the reversibility parameter. Seemingly, there are areas on the wafer where the energy of pinning sites is higher, which corresponds to the higher polarization reversibility, and is likely due to the inhomogeneous crystallization process caused by CMOS BEoL thermal budget restriction.

In future works, unsupervised machine learning methods like clustering can be harnessed to further clarify the dependencies and differences between device parameters with a full wafer-scale mapping. However, the trend observed in **Figure 3 c)** already demonstrates that the model parameters are not randomly assigned, and certain parameter dependencies exist, so that dimensionality reduction methods can be used to reduce the total number of parameters if operating conditions remain the same. However, a larger set of data for dimensionality reductions would provide a better result, which is a subject of future investigation.

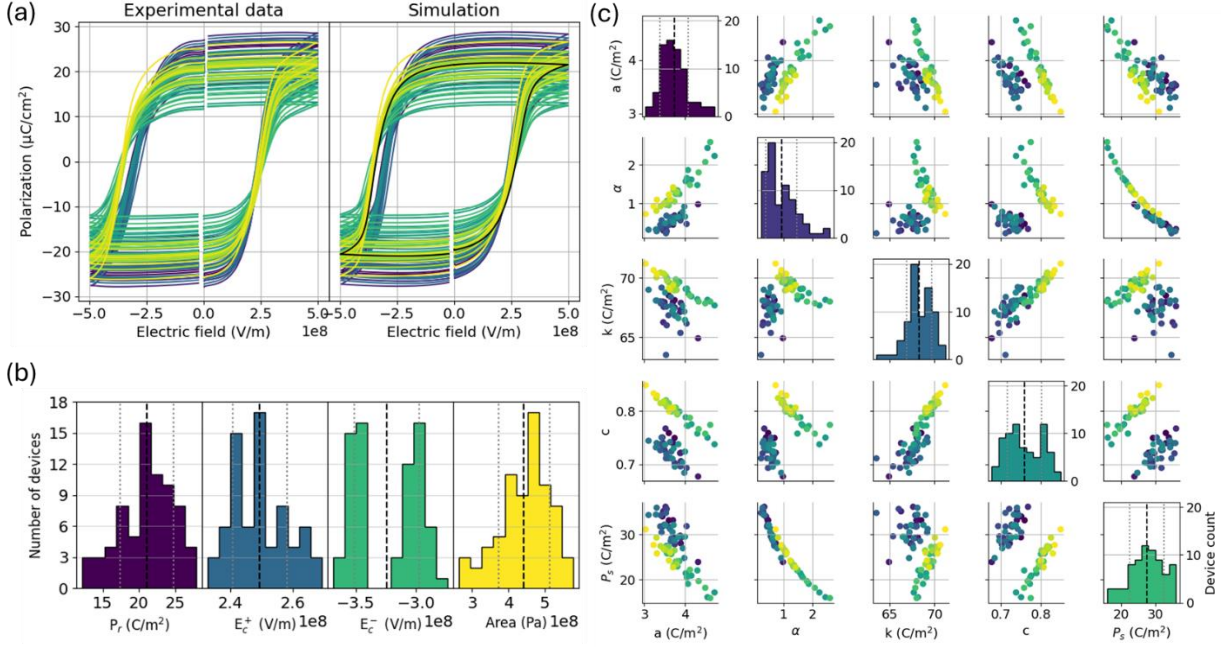

**Figure 3: a)** The $P - E$ curves for measured 73 devices showing device-to-device variation and the corresponding JA simulations. **b)** Statistical distributions of characteristics ($P_r$, $E_{C+}$, $E_{C-}$ and area) of each of the curves in **a)**. **c)** Parameter dependencies analyzed for the device-to-device variation analysis, with histograms showing the statistical distribution of the Jiles-Atherton parameter values on the diagonal. The colors and parameter values of each curve in **a)** and **c)** are shared.

## 2.2. Modelling of electric field dependent polarization switching

While polarization is often considered to encompass only the UP and DOWN states, the devices can also be operated in minor loop modes, where polarization is only partially switched while still exhibiting remanence. This is of utmost importance for analog memory components and for in-memory-computing (IMC). Microscopically, this partial switching corresponds to a fraction of the total number FE domains reversing direction. [36] The large memory window of hafnium-based FE devices allows us to distinguish these minor loops from each other for circuit applications, representing multilevel conduction states. [9] In **Figure 4,** we present results of modelling the partial polarization switching at a voltage range spanning from 7 to 10 $V_{pp}$ with a step of 0.5 volts. The results show that the JA model can accurately reproduce the experimental data for different applied voltages. The $P - E$ measurements and simulated data in the high voltage range reveal minor loop behavior is alike major loop, but with lower $P_r$ and $E_C$ values. Polarization switches partially when the input $E$ is not high enough to rotate the dipole moments of all the domains into alignment, corresponding to the ideal major loop, but strong enough to form a partially polarized stable state where the total polarization is still in alignment with the direction of $E$. Since the measured hysteresis curve changes under different operating voltages, the simulated model parameters also change with maximum input voltage. The minor loop behavior is especially important for circuit implementations with low power budget, where a switchable polarization is required with as low voltages as possible.

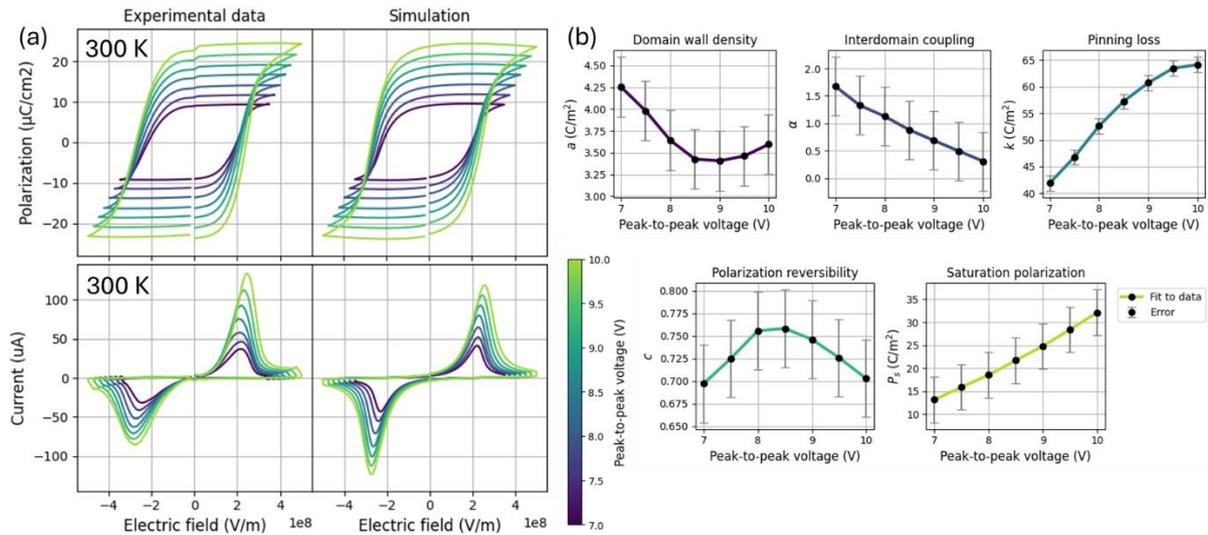

**Figure 4. a)** The experimental and simulated $P – E$ hysteresis loops with JA models. The JA loops show excellent agreement between experimental and model data from $V_{pp}$ = 10 V – 7 V. **b)** Applied voltage dependence of JA model parameters extracted from the experimental data. The error bars are formed from the standard deviation of each parameter in the presented device-to-device variation dataset. The lower voltage simulation results and data are shown in the supporting information (**Figure S2**)

The area under the $P – E$ loop not only gives an estimation of the energy loss over a cycle for a ferromagnetic/FE material for energy storage applications, but also reveals whether domain switching can be reproduced accurately from data with simulation. The accuracy of the simulated area compared to the experimental area under the $P – E$ loop reveal that the JA model shows an accuracy > 99 % in the 7 - 10 $V_{pp}$ range. Noteworthy, the JA model parameters do not stay constant under different operating voltages. This work presents how these parameters change under different voltage amplitudes. It contributes to both the computational understanding of the model for further device level simulations and provides a material level explanation of the behavior under different $E$. The change in the JA model parameters with respect to the $V_{pp}$ amplitude are presented in **Figure 4 b)**. These optimally fitted parameters describe the experimentally solved $P_r$, $E_C$ and the area for the voltage amplitude range from $V_{pp}$ = 10 V – 5.5 V with more than 98 % accuracy. In the voltage range of $V_{pp}$ = 5 V – 3 V (**Figure**

**S2**), the accuracies drop incrementally with the voltage amplitude to ~70 % due to the intrinsic stochasticity of the domain switching, as nucleation gains prominence over the domain wall movement. Below those voltages, only nucleation dominates the process, rendering the accuracy metrics obsolete with highly increased cycle-to-cycle variation on the barely open memory window and minimal remanent polarization.

The model accurately describes the measured imprint related asymmetry in the electric field axis with its fine-tunable inter-domain coupling, α and pinning loss parameters, *k*. The saturation polarization, $P_s$, is linearly correlated to the voltage amplitude, which is also linearly dependent on the measured $P_r$. With increasing $V_{pp}$, the values of α decrease as a function of $1/V_{pp}$. This is in accordance with the derivation by Smith et al.,[33] where the parameters are inversely dependent. Higher *E* decreases the effect of the domain coupling, which is regarded to the external *E* overpowering the interactions between neighboring domains at high fields. As α decreases with the increasing $P_s$, *k* increases with voltage amplitude. The almost linear relation between the voltage and pinning loss indicates that at higher voltages more pinning sites can be broken due to the higher energy in the domain walls. The polarization reversibility and domain wall density parameters show behavior alike each other, suggesting that the reversibility of the polarization is connected to the domain wall density. In a material where domain wall movement is more impeded, the polarization reversal requires higher effort than with a more unrestricted movement. The polarization reversibility first increases with voltage, possibly due to generation of new pinning sites with the cycling, but then decreases as these pinning sites get de-pinned with the added energy from higher *E*. The error accounted for D2D variation is relatively larger with these two parameters compared to the rest, which suggests that these two parameters are less affected by *E* than the rest of the parameters.

The hysteresis curves produced by the extracted parameters are found highly accurate to their experimental counterparts. Notably, these optimal parameters do not stay constant in different maximum *E* conditions. At higher voltages the energy is high enough to switch domains that are not switchable at lower voltages due to strong coupling between domains or domain pinning. Thus, the JA parameters for arbitrary pulse signal modelling should be chosen according to the highest voltage pulse, corresponding to a program pulse in memory applications. As this work presents, the JA model is accurate, efficient, and well suited for circuit simulations due to the continuous solution the model provides. The high accuracies obtained with the current computationally efficient model show that our model has potential for simulating the voltage dependent current switching in large-scale integrated FE devices in NVM circuits or for neuromorphic applications.

**2.3 Modelling of temperature dependent polarization switching**

The ferroelectric polarization of fluorites as a device-level characteristic has mainly been considered at room temperature, but as a microscopic physical effect, it is highly temperature dependent. As low-temperature applications like cold memories in data centers non-volatile memories in space applications, and memories for quantum computers are gaining significant research and industrial interest, the behavior of electronic components at low temperatures needs further research and development. It has been reported that at low temperatures, there is an increase in $P_r$, [19,20] while still upholding stability and endurance. [54] The low temperature operation of hafnium-based devices has not been simulated before, to the best of our knowledge. This work shows that JA model can accurately model FE polarization switching of the HZO devices down to deep cryogenic temperatures. Due to the domain kinetics-based derivation of the model, it can be used to assess changes in switching behavior at low temperatures considering the parameters at different temperatures.

The results of the temperature dependent simulation are presented in **Figure 5,** where the $P – E$ and dynamic $I – V$ curves measured and simulated at low temperatures are shown in **a)**, and the accuracies corresponding to the total temperature scale are represented by the colormap in **b)**. It is evident that the JA model can reproduce the $P – E$ curves highly accurately for $P_r$, $E_C$ values, and temperature alike, mostly in the 98-100 % range, and even the lowest accuracy observed was above 93 %. Slightly lower accuracies at low temperature can be explained by experimentally observed peak broadening into a double peak, which is not built-in to the assumptions of the JA model.

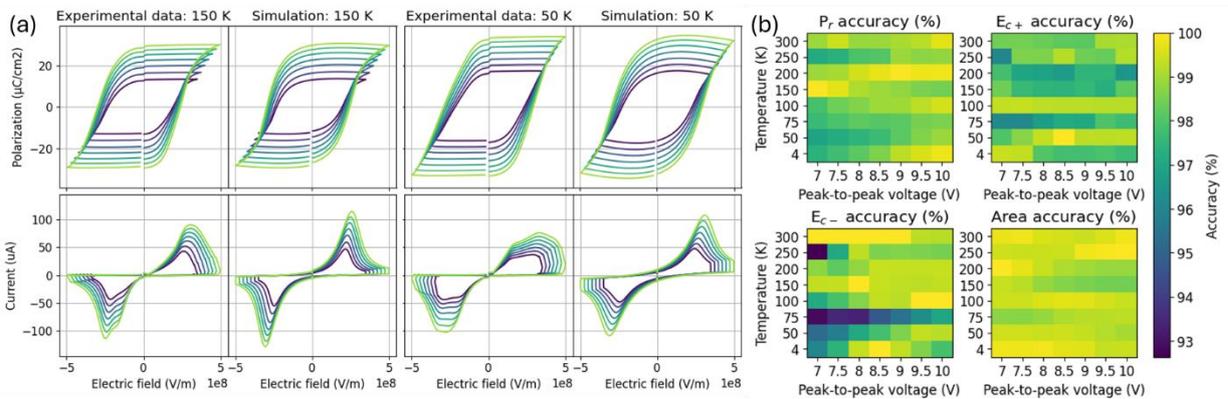

**Figure 5.** The experimental and simulated $P – E$ and dynamic $I – V$ curves at two different cryogenic temperatures of 200 K and 4 K, and the fitting accuracies corresponding to each of the temperature and voltages scales.

The parameter values for each of the measurements, mapping from 4 K – 300 K with a range of voltages from 7 – 10 $V_{pp}$, are presented in **Figure 6**. As the accuracies are high for each of these simulated curves, the parameter values reveal information on how the semi-microscopic domain wall kinetics change when the temperature is lowered. An increase in the polarization update linearity at or below 100 K was reported in our previous work, [19] and the parameter values obtained from the JA model also indicate that at 100 K and below, some operational difference occurs. For all the simulation parameters, the behavior above 100 K and below 100 K exhibit significantly differing results. We hypothesize that a low-temperature phase transition from non-polar to polar phase happens as the temperature lowers below this critical point, similarly to what was suggested by Xing et al., [20] where with first principles calculations, a tetragonal to orthorhombic phase transition is reported to be most favorable. This also explains the increase in experimental $P_r$ measured with the decrease in temperature, and the corresponding increase in the $P_S$ parameter of the JA model, as seen in **Figure 6**.

One temperature related phenomenon to consider is the thermal contraction when cooling down to deep cryogenic temperatures. In general, thermal contraction can be highly relevant in thin film devices, especially in devices with asymmetric electrode configurations, [55] like in this work. Modified strain in the FE material due to mismatch in thermal contraction between TiN, HZO and Au layers that can further influence the crystal phase transitions related to cooling. In addition, it can change the domain dynamics in the material, as described by Smith et al. [33], where the external stress is related to the inter-domain coupling parameter, α, with a lower stress corresponds to a lower value of α, which is also seen in **Figure 6**. The behavior observed with decreasing $a$ in **Figure 2** leads into a sharper current peak, and thus faster domain wall movement. As is seen from the temperature dependence of the parameter $a$, the velocity of the domain walls decreases at low temperatures, leading to more gradual switching, which corresponds to the widened current peaks in the experimental data. A suppression in thermally activated domain wall creep velocity has been experimentally observed in cryogenic FeFETs, [59] and is supported by the Merz law. [44] The parameter $k$ corresponds to the loss due to de-pinning of domains, and as is seen from **Figure 6**, the observed loss is decreased at low temperatures. Domain de-pinning occurs due to low energy charged defects, which are generated in charge injection with the applied electric field. The simulation results point to either a lower number of these charged defects being generated at low temperatures, or the activation energy required for a domain to de-pin is higher at lower temperatures, since observed pinning loss is lower at cryogenic temperatures or a mixture of both simultaneously.

The analysis in experimental work by Lancaster et al. [51] supports the assumption that the number of charged defects is lowered with decreasing temperature. Results by Bohuslavskyi et al., [19] suggests at lower temperature an improved $P_r$ value can be obtained with a high pulse voltage, confirming the need of higher energy to de-pin the domains. Charge trapping at cryogenic temperature needs to be further investigated experimentally to confirm these hypothesis more directly, since charge trapping is a major factor in device performance related to stability, latency and gradual or accumulative switching. [51] Both the loss related to de-pinning of domains, and the reversibility in **Figure 6** are decreased in low temperatures, supporting the assumption that in total a lower number of pinning sites is generated in low-temperature field cycling when compared to room temperature.

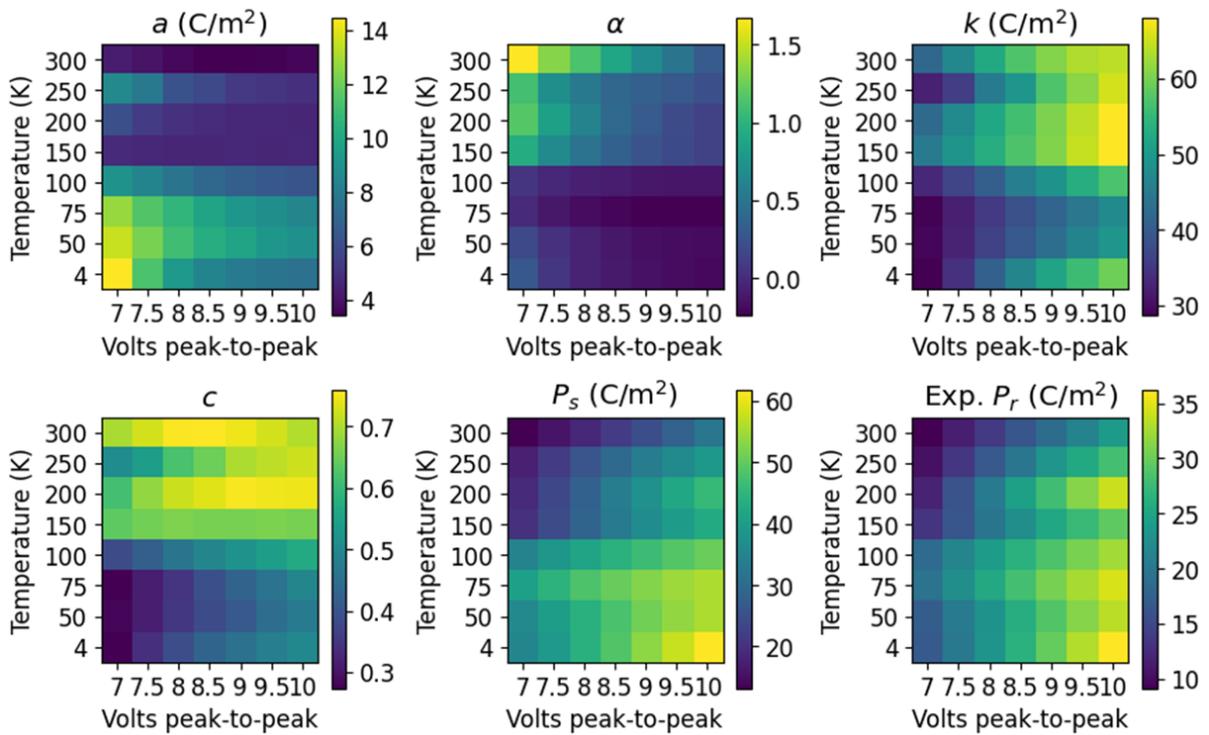

**Figure 6:** Parameter values as a mapping of temperature and applied voltage. Last panel is the experimental remanent polarization value extracted from the experimental *P-E* curves.

**2.4 Analog switching with subthreshold pulses**

As the JA model can simulate the voltage dependent polarization reversal, it can be used to predict the gradual switching of the polarization under different pulsing schemes. It provides valuable insights into the polarization update process, which can be harnessed to represent multiple analog memory states in NVM applications, such as for synaptic arrays for neuromorphic hardware accelerators, which highly benefit from symmetric and linear weight update process. [9,52] As reported by Bohuslavskyi et al., [19] cooling down to deep cryogenic temperatures causes an increase in the linearity of the polarization obtained with pulses over the coercive voltage. In this work, we simulate the remanent polarization update at different temperatures with two different subthreshold pulse schemes. According to the experimental Merz law, both domain nucleation and propagation are slower at low temperatures. [44] We propose that this slower nucleation and growth enables better controllability, enabling more symmetric and linear weight update protocols of the polarization update process with external electric field at lower temperatures.

In **Figure 7 a)**, the theoretical increase in polarization from 0 with a pulsing of identical 1 V pulses is shown. From this simulation, we find that the polarization increase process is slower at temperatures between 4 K – 75 K, than at and above 100 K. Simultaneously, the linearity of the polarization update is better at cryogenic temperatures. We assume that the increase in linearity is caused by the impeded domain wall growth and nucleation at lower temperatures. However, the clear separation between over 100 K and below 100 K, is most likely caused by the cryogenic tetragonal to orthorhombic phase transition discussed during the temperature dependent analysis and in work by Xing et al. [54]. Overall, this shows that there is an intrinsic difference in the polarization update at room and cryogenic temperatures and accordingly a customized protocol needs to be developed.

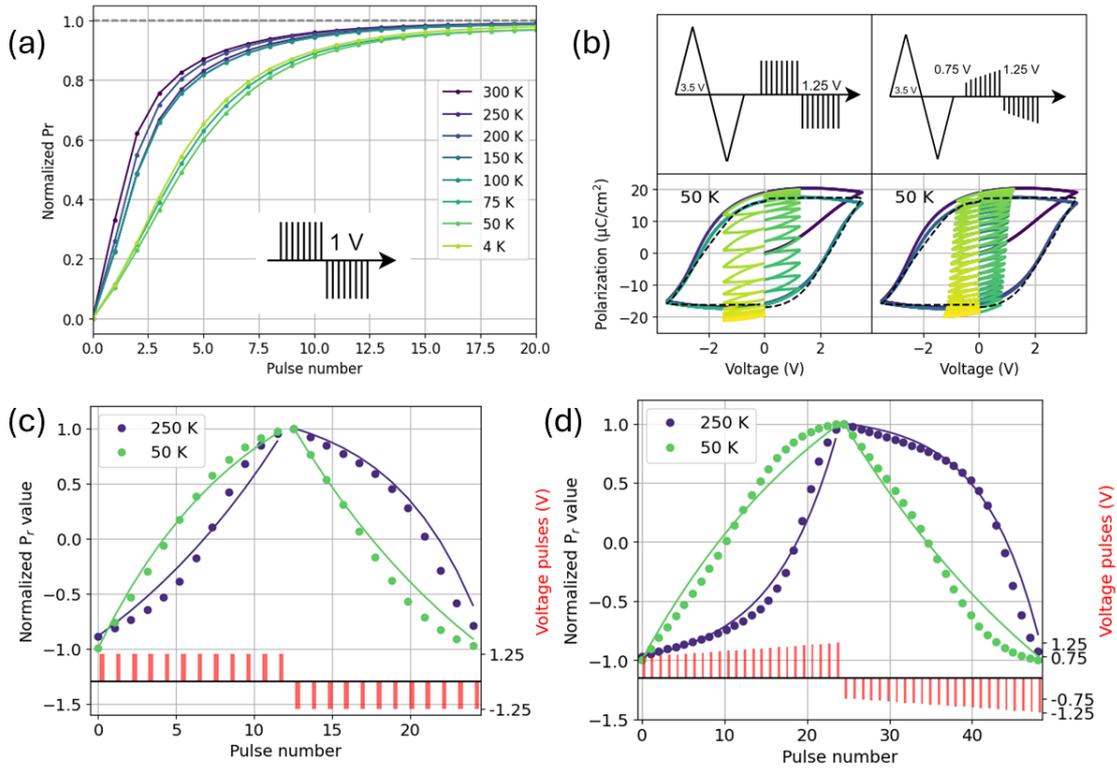

**Figure 7**: Simulated gradual switching of the polarization as a function of the input electric field and as the subthreshold pulse number. **a)** Simulated weight update from 0 polarization in different temperatures. **b)** JA solution to polarization electric field development pulsing of identical pulses, and for increasing pulse amplitudes, respectively. **c-d)** Simulated weight update including linearity with identical pulsing, and for increasing pulse amplitude, respectively. Potentiation (P) and depression (D) linearity coefficients were for the identical pulses P:-0.67, D:1.80 and P:1.05, D:-0.64, and for the increasing pulse amplitude P:-1.28, D:1.46 and P:0.39, D:-0.29 for 250 K and 50 K, respectively.

To better represent realistic device operation in simulations, the starting polarization state should be written with a larger write pulse. This is shown in the left panels in **Figure 7 b)**, where the initial polarization is written into $-P_r$ with 3.5 V triangular pulses, followed by subthreshold pulsing consisting of identical 1.25 V pulses. The corresponding potentiation and depression in the $P_r$ value is shown in **Figure 7c)** at both 250 K and in 50 K. The choice of the pulsing amplitude of 1.25 V is motivated by the fact that this pulse amplitude is not sufficient to switch the polarization rapidly, but domain nucleation and domain wall propagation can start at this pulse value and accumulate over multiple pulses to show gradual switching effect. Similar work using a transient Preisach model has been done previously, but only at room temperature. [9] The linearity factor is defined as in the Methods section, and the value 0 corresponds to a perfectly symmetric device. As seen from the figure and the linearity factors, there is a significant increase in linearity when cooling the device down. At low temperature, the symmetry of the polarization update is also increased, which is seen from the linearity factors approaching each other. Both at 250 K and 50 K, at least 8 intermediate polarization states can be distinguished with identical pulses, corresponding to programmed states in a 3-bit memory.

Finally, to demonstrate the maximal linearity, as was suggested in work by Jerry et. al, [9] we simulated the behavior with a pulse scheme of increasing pulses from 0.75 V up to 1.25 V, as is demonstrated in the right panels in **Figure 7b)** and in **Figure 7d)**. Custom pulse schemes require additional hardware, [52] however when custom pulse schemes are available for memory programming, highly linear and symmetric weight update behavior can be obtained. The simulations show that the linearity is further increased when cooling down to cryogenic temperatures, while the number of accessible polarization states is further increased with a custom pulse scheme. This pulse scheme also leads into highly symmetric potentiation and depression in cryogenic operation, with 24 accessible memory states, which can be used as 4-bit programmable memories.

It should be noted that the JA model does not explain stochasticity related to domain nucleation at very low voltages. The pulse amplitudes and times used for the pulsing simulations are such where the domain wall movement is more dominant than nucleation events, however, at room temperature the nucleation rates are so high that the domain wall movement under these small pulses cannot fully explain the analog nature of polarization switching with as small pulse amplitudes. At 250 K and below, the nucleation seems to slow down, and the full switching process can be explained with the JA model.

Even though the JA model can be used to model the polarization development under different pulsing schemes, there is no explicit frequency dependence in the model. This means that the pulse width is tied to the sample rate of the measurements, which is used as the time step in the model, controlling the time-development of polarization. The pulsing simulations were done with simulated 25 µs pulses, that are sinusoidally shaped to emulate physical smooth changes in voltage. Suggestions on implementing a frequency dependency to the ferroelectric polarization response into the Jiles-Atherton model through fractional differentiation, [60] but due to the complete history dependency of fractional derivatives, the SPICE and Verilog-A compatibility of the model would be lost in the process, so in this work the sample rate of measurements was used to tie to model time step to real time. The results obtained in this analysis show that HZO-based memory elements naturally exhibit higher linearity in the weight update process when cooled down to deep cryogenic temperatures, which is explained by both slower nucleation and impeded domain wall movement, enabling better controllability with applied voltage. This is an important result motivating further studies into the weight update process at cryogenic temperatures, where problems related to symmetry and linearity of the weight update process are mitigated through the reduced thermal energy, which allows for smaller energetic steps to be distinguishable.

## 2.5 Microscopic switching dynamics with Monte Carlo simulation

To further clarify the slower nucleation and domain wall movement at low temperatures, we performed a polycrystalline Monte Carlo simulation based on a dipole-dipole and dipole-electric field Hamiltonian, where the polycrystalline unit cells are simulated as dipole moments. Most Monte Carlo simulations of fluorite ferroelectrics are based on the statistical distribution of nucleation times in the NLS model. [46] On the contrary, this work approaches the polarization switching form a first principles point of view through the potential energy of the system, which is further explained in the Methods section. The simulations are done on a 50 x 50 x 20 grid, to emulate the physical thinness of the devices. Including the third dimension expands analysis both in the direction of the electric field called the out-of-plane direction, as well as the in-plane polarization, both of which corresponds to one layer of unit cells. To include the spatial constrains induced by the polycrystalline structure, a q-state Potts model [61] is used to stochastically obtain the spatial variations in the polycrystalline structure, where only parts of the total area of the device correspond to the ferroelectric orthorhombic phase.

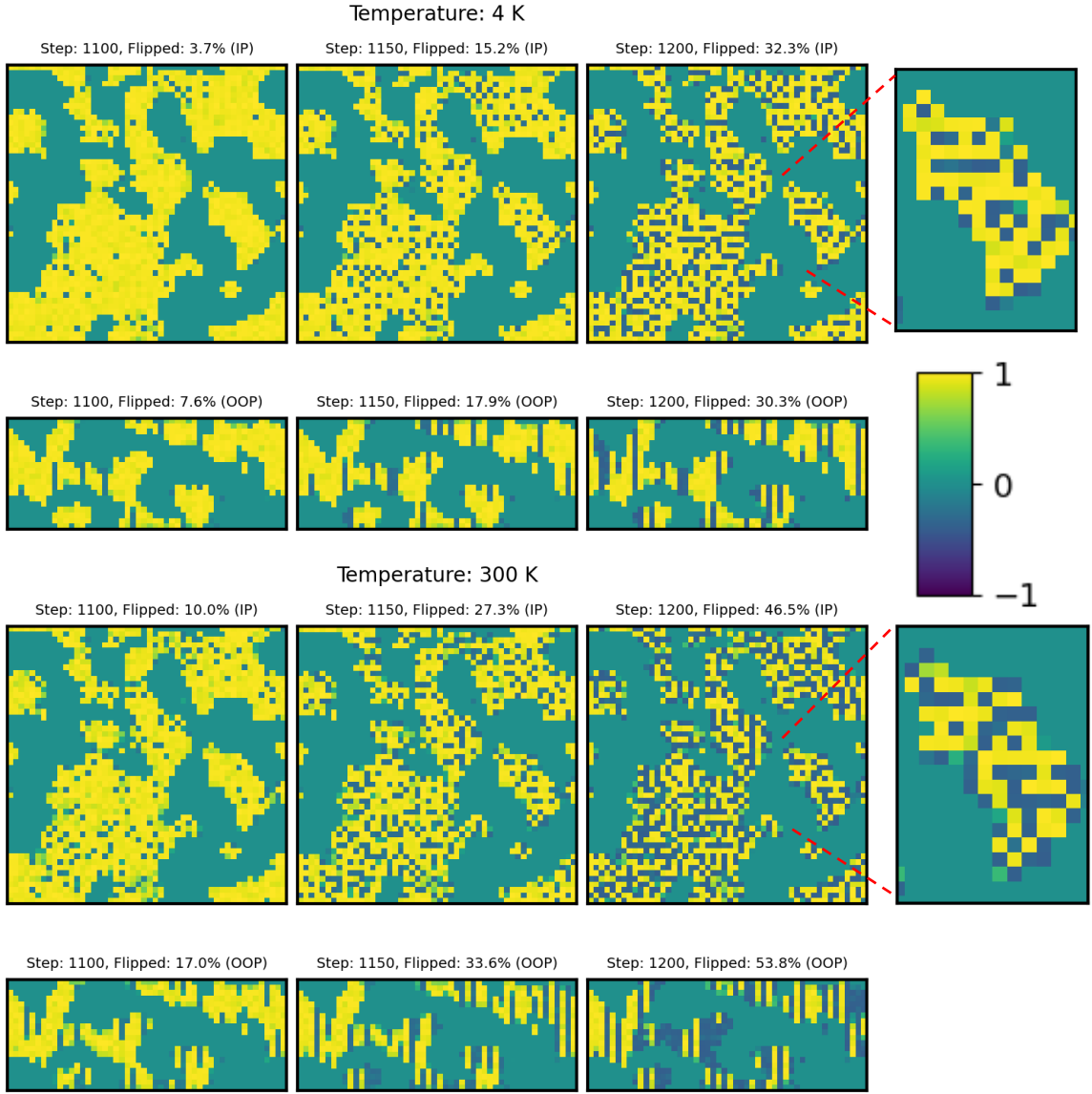

**Figure 8:** Microscopic Monte Carlo simulation of ferroelectric polarization switching based on dipole interactions at three distinct time steps for both temperatures. Both the in-plane (IP) and out-of-plane (OOP) directions are shown. Each grid point represents a unit cell, where the cyan depicts the non-orthorhombic phases where polarization is always zero, and yellow-to-blue corresponds to the opposite polarization states with intermediary steps in between, indicated by the normalized color bar. The percentage indicated corresponds to the ratio of cells flipped from $P_s$ to -$P_s$.

The microscopic polarization development is shown if **Figure 8**, where each grid point corresponds to a unit cell with a dipole moment, and the total polarization is a sum of the out-of-plane components of these dipole moments. In these simulations, nucleation happens more easily at defect sites or grain boundaries, due to these sites having the lowest number of orthorhombic nearest neighbors, and most nucleation happens right next to the electrode interfaces. At low temperatures the nucleation sites form at the edges of the FE material, from which it rapidly permeates in the out-of-plane direction, without much growth in the in-plane direction. With rising temperatures, the nucleation happens faster, and the domains also slightly permeate in the in-plane direction, which is seen as faster polarization switching. This vertical over lateral growth observed in these Monte Carlo simulations is what the experimental Merz law states [44], down to the temperature dependency. Due to the limited dimensionality in the out-of-plane direction of the thin films, the Merz law reduces into the NLS model. These vertical growth patterns form head-to-tail paths of dipole moments, and a singular nucleation does not cause the full polarization switching of the system, which explains the existence of multiple incremental polarization states.

The total polarization in the simulation was validated with the same set of experimental data, and the simulated hysteresis curves are in good agreement with both the experimental results and the JA model. Through this Monte Carlo analysis, the domain wall movement in the out-of-plane direction follows the assumptions of the JA model. At very small voltages (<0.75 V) at room temperature, where nucleation events dominate, the JA framework is not able to explain the full polarization development like the Monte Carlo simulation. However, at low temperatures, where nucleation events are scarcer and domain wall movement happens less swiftly, the domain wall dynamics do follow the JA assumptions at lower voltages than at high temperatures.

Simulations with a similar Hamiltonian was done in work by Lee et al. [62], where the in-plane nucleation was investigated. The analysis was limited to a two-dimensional system, which removes the possibility of domain wall movement in the out-of-plane direction. Since no polycrystalline structure was implemented, the simulations corresponded to an epitaxial system, where the dynamics of the KAI model are followed, but a more NLS-like switching is obtained with the addition of defects. Expanding on these results, we note that to obtain the complete polarization switching behavior, the complete three-dimensional degree of freedom is required to see the vertical domain growth in addition to the nucleation. We present that through initializing the three-dimensional polycrystalline structure, the vertical domain growth of the polarization development and stable partially switched polarization states can be visualized with results in agreement with the NLS model and Merz law.

While the Hamiltonian used in the model can produce the vertical domain growth patterns, it can be expanded in future work to encompass other polarization related phenomena, such as interface energies and surface pressure of the domain walls. As of now, it is purely based on the potential energies of dipole interactions. While the simulated thickness of the FE material has been fixed to represent the 10 nm HZO devices, the area of the material in the simulations does not correspond to the size of the fabricated devices, due to the computational time complexity of the Monte Carlo model. While larger systems yield smoother polarization switching development, the operational logic is already seen with the smaller simulated systems. The Monte Carlo modelling done in this work confirms that the assumptions of the NLS model are naturally obtained from the head-to-tail dipole formation in spatially restricted domain growth environments, and that the NLS and JA models do not contradict each other but are relevant at different temporal operation regimes. The simulation results show that both nucleation rate and domain wall movement are slowed down at cryogenic temperatures. This further points at a higher controllability of the analog polarization states obtainable at cryogenic temperatures with pulsing.

## 3. Conclusions

In conclusion, we presented a physics-based SPICE compatible model that can predict with nearly 100 % accuracy the remanent polarization, coercive fields and area inside the *P-E* curve under different operating voltages down to deep cryogenic temperatures. We used these model parameters to simulate the polarization update process under different pulsing schemes both close to room temperature and down to deep cryogenic temperature. The results show that cooling ferroelectric devices down to deep cryogenic temperatures makes the polarization update process more linear and symmetric. We then further show that the analysis from the simulation results can also be obtained through a Monte-Carlo approach with a simple dipole interaction-based Hamiltonian. The out-of-plane domain growth produced with the Monte-Carlo simulations also clearly obey Merz law and the NLS model.

## 4. Methods

**Experimental: Fabrication:** The TiN/HZO/Al$_2$O$_3$/Au capacitors were fabricated with the bottom electrode (TiN) being fabricated using PEALD and the HZO layer using thermal atomic layer deposition (ALD). TiN thickness was of 30 nm and HZO film thickness of ~10 nm is reported in this study. The TiN bottom electrode (BE) was fabricated on Si/SiO$_2$ wafers with 500 nm thermal oxide on Si, using PEALD. Fabrication temperature of 420°C and TiCl$_4$ and ammonia were used as precursors for TiN growth, where nitrogen was used as carrier and purging gas, and argon flow was used for the plasma. HZO film was deposited at 200 $^0$C on with tetrakis (dimethylamino)hafnium (TDMAH) and tetrakis (dimethylamino)zirconium (TDMAZ) as the Hf and Zr precursors, respectively, and water (H$_2$O) as the oxidant. Nominally optimized 50:50 ratio of the Hf:Zr was obtained using one cycle of TDMAH and H$_2$O followed by a cycle of TDMAZ and H$_2$O. The thickness of the TiN layer was done with cross-sectional SEM while HZO thickness was measured using profilometry. Thicknesses of the film varied slightly from the targeted thickness. For instance, 1200 cycles of TiN resulted in thicknesses of 27 nm and 54 supercycles of HZO resulted in nearly 9.3 nm HZO. Right after the deposition of HZO, 1.2 nm Al$_2$O$_3$ was deposited at 200 $^0$C. Following the film deposition, rapid thermal annealing (RTA) step was done for 30 sec at annealing temperature of 500 $^0$C under the nitrogen atmosphere. After RTA, Au top electrodes were deposited as the top electrode via thermal evaporation device to complete the capacitors with different sizes of 100 μm × 100 μm, 200 μm × 200 μm and 500 μm × 500 μm.

**Characterization:** All electrical characterizations were done using Keysight's B1530A waveform generator outputting various waveforms and triggering a digitizing oscilloscope for fast data acquisition and a commercial transimpedance amplifier (*I-V* converter). A cryogenic probe station is used for varying temperatures between 4 K – 300 K. The data were post-processed by smoothing with a Savitzky-Golay filter to reduce noise without distorting the data. Details of the experiments are described in earlier publication. [19]

**Modelling:** The HZO capacitor devices are simulated with the computationally efficient semi-empirical Jiles-Atherton model based on five physically motivated fitting parameters. The electric fields and polarization inside the ferroelectric capacitor can be viewed as scalar quantities due to the electric field inside the dielectric material being highly homogeneous.

The depolarization field, $E_{\text{dep}}$, is included into the total electric field inside the material, $E_{\text{tot}}$, at each time point as suggested in [38]

$$E_{tot} = E_{bias} - E_{dep} = E_{bias} - \frac{P}{\epsilon_0 \epsilon_r}, \qquad (1)$$

where $E_{\text{bias}}$ corresponds to the electric field caused by the external bias, $P$ is the total polarization causing the depolarization field, and $\epsilon_0$ and $\epsilon_r$ correspond to vacuum permittivity and relative permittivity of the FE material, respectively.

The ideal (anhysteretic) polarization is shaped like the Langevin equation [33], obtained from assuming that the rotation of dipole moments is not restricted, and instead the dipole moments can rotate in all spatial directions when an external electric field is present. For noninteracting moments the total anhysteretic polarization, $P_{\text{anh}}$, is

$$P_{anh} = P_s \left[ coth\left(\frac{E_{tot}}{a}\right) - \frac{a}{E_{tot}} \right] \qquad (2)$$

where $P_s$ is the macroscopic parameter describing the saturated anhysteretic polarization, and $a$ is a loss term connected to the density of domain walls prohibiting the domain growth occurring due to the electric field.

At the macroscopic scale, the total electric field induces movement of free charges, and thus the displacement field $D$

$$D = \epsilon_0 E_{tot} + P, \tag{3}$$

where $\epsilon_0$ is vacuum permittivity, should be considered instead of the electric field to include the dielectric contribution to the displacement field. However, domains, which are effective electric dipoles, have interactions with each other that can be included by adding an interaction term, forming an effective displacement field $D_{eff}$,

$$D_{eff} = D + \alpha P \tag{4}$$

where $\alpha$ is the inter-domain coupling parameter. Only the interactions between the field and the irreversible polarization contribution $P_{irr}$ should be included due to localization of the coupling field [33]. This effective displacement can further be influenced by effective polarization due to applied strain, and thus the parameter $\alpha$ is dependent on the external stress conditions of the ferroelectric material [33]. The final form of the equation used in the simulations is then

$$P_{anh} = P_s \left[ \coth\left(\frac{D_{eff}}{a}\right) - \frac{a}{D_{eff}} \right]. \tag{5}$$

The energy related to the irreversible polarization is the difference between the ideal anhysteretic polarization energy and energy to break pinning sites, $\mathcal{E}_{pin}$, which are described by a pinning loss parameter $k$ as

$$\mathcal{E}_{pin} = k \int_0^P dP. \tag{6}$$

Being linearly dependent on both the average energy to break a singular pinning site, and the number of pinning sites to be broken, it contains information on the microstructure of the ferroelectric material. By considering the work done by the domain wall pressure to bend of domain walls as well as domain wall movement, which corresponds to domain growth, a differential equation governing the domain wall dynamics is obtained,

$$\frac{dP_{irr}}{dE} = \frac{P_{an} - P_{irr}}{\delta k - \alpha(P_{anh} - P_{irr})} \tag{7}$$

where $\delta = \text{sign}(dD)$ is included to make sure that breaking pinning sites requires energy instead of providing the system with more energy. Under nearly quasistatic conditions, the total polarization is a sum of the ideal anhysteretic and the irreversible polarization with a reversibility factor, $c$,

$$P(t) = cP_{anh}(t) + (1-c)P_{irr}. \tag{8}$$

To solve the ordinary differential equation for the change of irreversible polarization, the Euler integrator method for numerical integration is employed as,

$$P_{irr}(t) = P_{irr}(t - \Delta t) + \frac{dP_{irr}(t)}{dt} \cdot \Delta t, \quad (9)$$

Extracting the parameters from the experimental data was done with the curve fitting tools provided by SciPy in Python 3. The current can be solved as the time derivative of the polarization, $I_{pol} = \frac{dP(t)}{dt}$, which can be numerically obtained from the simulated polarization. In the pulsing simulations, the pulse width was 25 µs, and the integration time step was equal to that of the sampling rate in the experimental data, about 500 ns.

The linearity coefficient used to characterize the pulsing simulations, following the work of Chen et al. [52], is defined as $v$ in the linearity equations for potentiation, $P_{pot}$, and depression, $P_{dep}$,

$$\begin{cases} P_{pot} = B \cdot (1 - e^{-p \cdot v/10}) + P_{min} \\ P_{dep} = -B \cdot (1 - e^{(p - p_{max}) \cdot v/10}) + P_{max} \end{cases} \quad (10)$$

where $B$ is a function depending on the switching characteristics

$$B = \frac{P_{max} - P_{min}}{1 - e^{-p_{max} \cdot v/10}}. \quad (11)$$

Here $P_{max}$ ($P_{min}$) corresponds to the highest (lowest) obtainable polarization value, $p$ represents the variable of pulse number, $p_{max}$ the maximum pulse number required to change the polarization.

The model for the Monte Carlo simulations is based on a dipole approximation, where each orthorhombic unit cell is replaced with a dipole with three-dimensional rotational degrees-of-freedom. These dipoles stay localized within their unit cells and only rotate and strengthen with respect to the electric field as well as with respect to the nearest neighboring dipole unit cells. As the dipoles of the system are at rest within their unit cells, the canonical momentum is zero and the Hamiltonian of the system is obtained as the total potential energy, which is the sum of two dipole interactions: inter-dipole ($U_{\text{dipole – dipole}}$) and dipole-field ($U_{\text{dipole – field}}$) potential energies.

$$\text{H} = U_{\text{dipole – dipole}} + U_{\text{dipole – field}}.$$

This expression can be obtained from basic electrostatics with a nearest neighbor approximation into the form

$$\text{H} = \frac{1}{2} \sum_i \sum_{\langle i,j \rangle} \frac{\mu_i \mu_j}{4\pi\varepsilon r^3} (\gamma(\hat{p}_i \cdot \hat{p}_j) - 3(\hat{p}_i \cdot \hat{r})(\hat{p}_j \cdot \hat{r})) - \sum_i \frac{\alpha \mu_i \phi}{d} (\hat{p}_i \cdot \hat{E})$$

where $\hat{p}_i$ denotes the dipole unit vector corresponding to location $i$, with $\mu_i$ as the magnitude of said dipole. The vector $\hat{r}$ is the distance unit vector between the dipoles $\hat{p}_i$ and $\hat{p}_j$, $r$ is the distance between the two dipoles, ϕ is the potential difference and $d$ is the distance between the plates of the capacitor. The variable $\hat{E}$ denotes the directional unit vector of the electric field as it is defined as homogeneous within the capacitor. The summation indices $\langle i, j \rangle$ denote the nearest neighboring pairs, and scaling factor of ½ is needed to remove permutations of the pairs. The parameters $\alpha$ and $\gamma$ are implemented as multiplicative parameters to match the results better with the measured values.

The magnitude $\mu_i$ is voltage dependent, allowing for stable levels of polarization with varying fields, which is implemented as a linear function scaled with respect to $\hat{E}_{max}$. This subsequently causes the parameter $\alpha$ to be necessary, as the first term of the Hamiltonian scales with $\mu^2$ and the latter term with the factor $\mu$ requiring an additional multiplicative term relating to $\hat{E}_{max}$ as the field itself is not strong enough to flip the states. The parameter $\gamma$ weakens the anti-ferroelectric ordering, which in this model was set to ¾, to match the coercive field of the hafnia-based devices. The total effective polarization is obtained as sum of the $\hat{z}$-directional components of all dipoles.

$$P = \frac{1}{N} \sum_i \hat{p}_i \cdot \hat{z}$$
$$= \frac{1}{N} \sum_i \mu_i \cos\theta_i$$

To implement a temperature dependence into the system, the Metropolis-Hastings selection rule was imposed. This selection rule defines a cost function for the system, with respect to which the system will minimize itself. For all physical systems and subsequently this model, the cost function implemented is the Hamiltonian, which approaches a local minimum. This is done by calculating the energy with the Hamiltonian, causing a perturbation within the system and re-evaluating the energy. The Metropolis-Hastings selection rule returns the probability $P$ of the new, perturbed state to be accepted, as defined by

$$P(\Delta H) = \begin{cases} e^{-\frac{\Delta H}{k_B T}}, & \text{if } \Delta H > 0 \\ 1, & \text{if } \Delta H \leq 0 \end{cases},$$

where $\Delta H$ is the energy difference caused by the perturbation, $k_B$ is Boltzmann constant and $T$ the temperature.

The new, perturbed state will always be accepted, in cases where the total energy of the system decreases. In a suitable temperature for the energy difference, the new state can still be accepted, forming a basis for the temperature dependence of the system. With the Metropolis-Hastings selection rule, the system tries to approach a state which minimizes the energy through sampling the Hamiltonian of the system at each step.

**Acknowledgements**

The authors acknowledge financial support from Research Council of Finland through projects AI4AI (no. 350667), Ferrari (no. 359047), Business Finland and European Commission through project ARCTIC (no. 101139908). Authors acknowledge Heorhii Bohuslavskyi of VTT Technical Research Centre of Finland for providing the original experimental data files.

# Supporting Information

**Universal Model for Ferroelectric Capacitors Operating Down to Deep Cryogenic Temperatures**

*Ella Paasio, Rikhard Ranta, Sayani Majumdar\**

Faculty of Information Technology and Communication Sciences, Tampere University, 33720 Tampere, Finland

E-mail: sayani.majumdar@tuni.fi

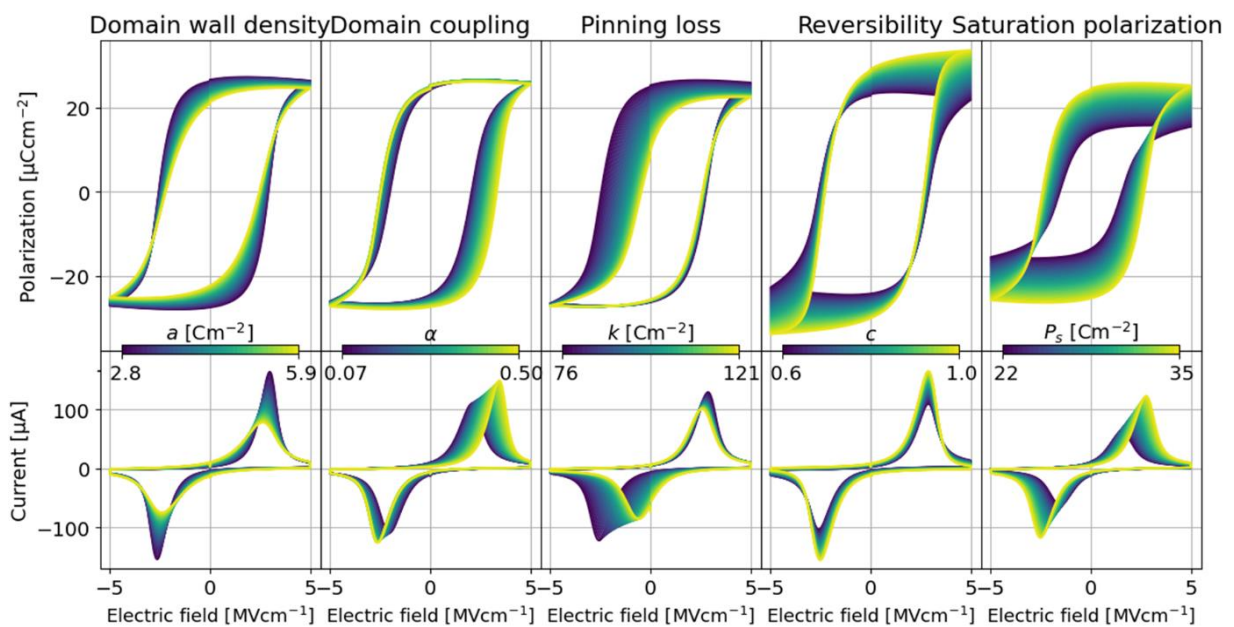

**Figure S1.** Changes in the hysteretic *P-E* curves with modifying one JA model parameter while keeping others constant and the corresponding *I-E* curves. The polarization was initialized with a negative voltage pulse, followed by a positive one. The color indicates the value of the modified parameter.

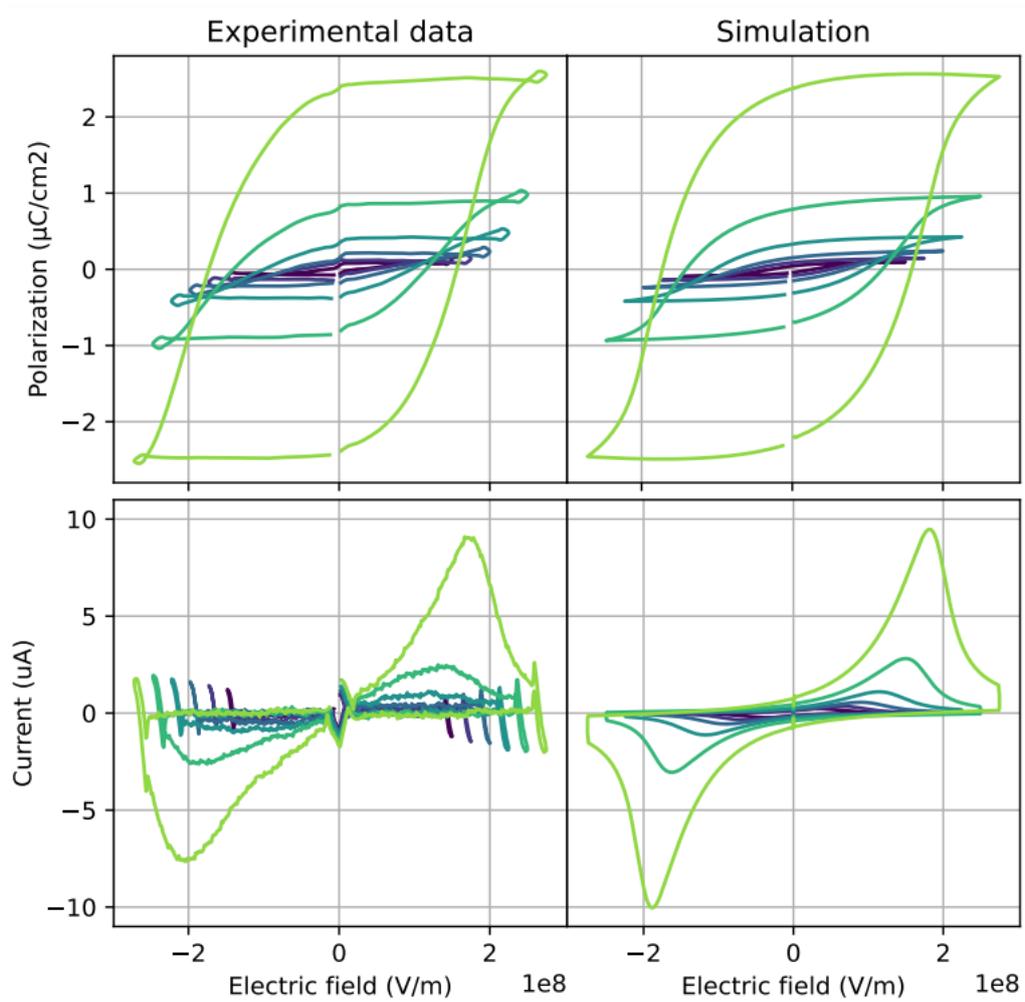

**Figure S2.** The experimental and simulated *P – E* and dynamic *I – V* curves with a full range of voltages on the right and only the low voltages in the left.